\documentclass[preprint,amsmath,amssymb,nofootinbib]{revtex4-1}
\usepackage{color}
\usepackage{graphicx,epstopdf}
\usepackage{dcolumn}
\usepackage{bm}
\usepackage[ansinew]{inputenc}
\begin{document}

\begin{abstract}
We study sterile neutrinos in an extension of the standard model, based on the gauge group $SU(3)_C\otimes SU(3)_L\otimes SU(3)_R\otimes U(1)_N$, and use this model to illustrate how to apply cosmological limits to thermalized particles that decouple while relativistic. These neutrinos, $N_{aL}$, can be dark matter candidates, with a keV mass range arising rather naturally in this model. We analyse the cosmological limits imposed by $N_{eff}$ and dark matter abundance on these neutrinos. Assuming that these neutrinos have roughly equal masses and are not CDM, we conclude that the $N_{eff}$ experimental value can be satisfied in some cases and the abundance constraint implies that these neutrinos are hot dark matter. With this information, we give upper bounds on the Yukawa coupling between the sterile neutrinos and a scalar field, the possible values of the VEV of this scalar field and lower bounds to the mass of one gauge boson of the model.
\end{abstract}

\title{Cosmological bounds of sterile neutrinos in a $SU(3)_C\otimes SU(3)_L\otimes SU(3)_R\otimes U(1)_N$ model as dark matter candidates}

\author{Cesar P. Ferreira}
\email{cesarpf@ifi.unicamp.br}
\author{Marcelo M. Guzzo}
\email{guzzo@ifi.unicamp.br}
\author{Pedro C. de Holanda}
\email{holanda@ifi.unicamp.br}
\affiliation{Instituto de Física Gleb Wataghin - UNICAMP, 13083-970 Campinas SP, Brazil.}

\maketitle

\section{Introduction}

Despite its effectiveness, the Standard Model of Particle Physics remains unable to solve some important problems, among them, the Dark Matter (DM) problem. This is a problem related to discoveries made by astronomical observations, and DM play a central role in the standard cosmological model (known as $\Lambda$-CDM). Dark Matter probably is a kind of particle, as yet unknown, that certainly interacts trough gravity, maybe interacts trough the weak force, and doesn't interact trough electromagnetism and the strong nuclear force. It represents $~ 27\%$ of the energy content of the Universe.

This problem is among the major motivations for the study of extensions of the SM. However, any candidates to dark matter particles must satisfy constraints imposed by cosmology. If the model predicts somewhat light candidates (keV range or lower), these can be Warm or Hot Dark Matter, with the former possibility being the most attractive one, since the latter is restricted to have at most a few percent of the energy density of Cold Dark Matter. If these light candidates are thermalized, due to exchange of gauge bosons with SM particles, then its number density has a fixed relation with its temperature. With this in mind, its possible to impose the $N_{eff}$ (related to radiation energy) and abundance constraints to any particle satisfying these requirements. Not only that, but limits on the mass of the gauge boson that mediates the interaction can also be given, which constrain the value of the VEV (Vacuum Expectation Value) that is related to this gauge boson mass and the mass of any other particle that depends on this VEV.

Several models have viable candidates for DM. In Ref \cite{331LR}, a model was proposed that attacks the DM problem, based on the gauge group $SU(3)_C\otimes SU(3)_L\otimes SU(3)_R\otimes U(1)_N$ (the 3L3R model for short). This model modifies the electroweak sector of the SM, making the substitution $SU(2)_L\otimes U(1)_Q\rightarrow SU(3)_L\otimes U(1)_N$, that usually defines the 3-3-1 models\cite{Pis:1991,Dias:2006} and adds a $SU(3)_R$ group to obtain a Left-Right symmetry at higher energies. The neutrino sector is an important part of this model, and in it the model predicts the existence of 12 neutrinos: 3 of them are eV-range neutrinos, identified as the usual active neutrinos; 3 are left-handed,  keV-range sterile neutrinos, $N_{aL}$, candidates for Warm Dark Matter (WDM), and the remaining 6 are right-handed, ultra heavy neutrinos, with masses around $10^{11}$ GeV, that could play a role in baryogeneses through leptogenesis in the primordial universe. These neutrinos will be in this mass range, provided that the Yukawa coupling constants between these neutrinos and the scalar fields of the model are of $\mathcal{O}(1)$.

Our main objective in this paper is to analyse the viability of the keV neutrinos of the 3L3R model as candidates for dark matter, and use this model as an example to particle physicists of how to apply cosmological bounds to light, stable particles. We impose the known restrictions in $N_{eff}$ and DM abundance to several cases: (1) Decoupling of these neutrinos before $\mu,\mu^+$ annihilation; (2) Decoupling before pion annihilation; (3) Decoupling before hadronization. These different decoupling temperatures also give lower bounds on the mass of the gauge boson, $U_L$, that mediates interactions between $N_{aL}$ and charged leptons in the primordial plasma. In this analysis, a critical assumption is made: That all left-handed sterile neutrinos have roughly equal masses and are not CDM, which in essence ensures that they are stable and that they decouple while relativistic. With this, we reproduce a known result that thermalized keV stable particles are overproduced, so we must conclude that the new left-handed neutrinos are Hot Dark Matter. Finally, restrictions on the Yukawa couplings of these neutrinos and a scalar field are given.

The paper is organized as follows: In section 2, the 3L3R model is presented. In section 3, the sterile neutrinos temperature and abundance are determined. Section 4 gives the $N_{eff}$ of the sterile neutrinos. In section 5, the decoupling temperature of the sterile neutrinos is determined as a function of the mass of the $U_L$ gauge boson. Section 6 gives our results, and Section 7 is reserved to our conclusions.

\section{The 3L3R Model}

As mentioned earlier, the 3L3R model is based on the gauge group $SU(3)_C\otimes SU(3)_L\otimes SU(3)_R\otimes U(1)_N$. With this group, the electric charge operator is given by\cite{331LR}
\begin{equation}\label{q2}
Q=T^{3}_{L}+T^{3}_{R}-b\left(T^{8}_{L}+T^{8}_{R}\right)+N,
\end{equation}
where $T^{3}_{L,R}=\lambda^{3}/2$ and $T^{8}_{L,R}=\lambda^{8}/2$ are the diagonal generators of the $SU(3)_{L,R}$ groups, N is the generator of the $U(1)_N$ group, and $b$ is a real parameter. The $\lambda^{i}$ are the Gell-Mann matrices. One nice consequence of the 3L3R is that the requirement that the fermions have integer electric charges, combined with the restriction that the Weinberg angle, $\theta_W$, satisfies $sin^2\theta_W=0.231$\cite{PDG} implies that $b=1/\sqrt{3}$\cite{331LR}.

The leptons in the model are organized in triplets of the $SU(3)_{L,R}$ groups,
\begin{eqnarray}\label{3l3r.trip.lep}
\Psi_{aL}&=&\left(\nu_{aL}, l_{aL}, N_{aL}\right)^T\sim \left({\bf 1}, {\bf 3}, {\bf 1}, -1/3\right),\\
\Psi_{aR}&=&\left(\nu_{aR}, l_{aR}, N_{aR}\right)^T\sim \left({\bf 1}, {\bf 1}, {\bf 3}, -1/3\right),
\end{eqnarray}
where $a=e,\mu,\tau$ are the three leptonic families, and $N_a$ are new neutrinos. L and R indicate Left and Right components of the fields. The numbers in parenthesis indicate how these particles transform under the $SU(3)_C$, $SU(3)_L$, $SU(3)_R$ and $U(1)_N$, respectively. 

The scalar sector consists of 3 $SU(3)_L$ and 3 $SU(3)_R$ triplets,
{\small
\begin{equation}\nonumber
	\begin{array}{l}
\eta_{L}=\left(\eta_{L}^{0},\eta^{-}_{L},\eta^{'0}_{L}\right)^T\sim \left({\bf 1}, {\bf 3}, {\bf 1}, -1/3\right),\\
 \eta_{R}=\left(\eta_{R}^{0},\eta^{-}_{R},\eta^{'0}_{R}\right)^T\sim \left({\bf 1}, {\bf 1}, {\bf 3}, -1/3\right),\\ \nonumber
\rho_{L}=\left(\rho_{L}^{+},\rho^{0}_{L},\rho^{'+}_{L}\right)^T\sim \left({\bf 1}, {\bf 3}, {\bf 1}, 2/3\right),\\
 \rho_{R}=\left(\rho_{R}^{+},\rho^{0}_{R},\rho^{'+}_{R}\right)^T\sim \left({\bf 1}, {\bf 1}, {\bf 3}, 2/3\right),\\ \nonumber
\chi_{L}=\left(\chi_{L}^{0},\chi^{-}_{L},\chi^{'0}_{L}\right)^T\sim \left({\bf 1}, {\bf 3}, {\bf 1}, -1/3\right),\\
 \chi_{R}=\left(\chi_{R}^{0},\chi^{-}_{R},\chi^{'0}_{R}\right)^T\sim \left({\bf 1}, {\bf 1}, {\bf 3}, -1/3\right).
	\end{array}
\end{equation}
}
Only 6 of the 10 neutral components, $\chi^{'0}_{L,R}, \eta_{L,R}^{0}, \rho^{0}_{L,R}$, have a non-vanishing Vacuum Expectation Value(VEV), that are given by,

\begin{equation}\nonumber
	\begin{array}{ll}
\left<\eta_{L}^{0}\right>=\nu_{\eta_{L}}/\sqrt{2}, & \hspace{1cm} \left<\eta_{R}^{0}\right>=\nu_{\eta_{R}}/\sqrt{2},\\ 
\left<\rho_{L}^{0}\right>=\nu_{\rho_{L}}/\sqrt{2}, & \hspace{1cm} \left<\rho_{R}^{0}\right>=\nu_{\rho_{R}}/\sqrt{2},\\\nonumber
\left<\chi_{L}^{'0}\right>=\nu_{\chi_{L}^{'}}/\sqrt{2}, & \hspace{1cm} \left<\chi_{R}^{'0}\right>=\nu_{\chi_{R}^{'}}/\sqrt{2}.
	\end{array}
\end{equation}

Since the Left-Right symmetry happens only at very high energies, it is assumed that the VEV's associated with the $SU(3)_R$ scalar triplets are much higher than those associated with the $SU(3)_L$ triplets, that is to say, $\nu_R>>\nu_L$. And, since new particles  in the left triplets gain masses due to the $\nu_{\chi_L^{'}}$ VEV, it is imposed that $\nu_{\chi_L^{'}}>>\nu_{\eta_L},\nu_{\rho_L}$. Like most 3-3-1 models, it is assumed that $\nu_{\chi_L^{'}}$ is in the TeV range. This happens to keep these new particles heavier than the known quarks and weak gauge bosons. 

In the gauge sector, there exists 16 weak gauge bosons. The known $W^{\pm}_L, Z_L$ of the SM, $Z^{'}_L, U^{\pm}_L, V^0_L, (V^0)_L^{*}$ are new left handed gauge bosons, and $W^{\pm}_R, Z_R, Z^{'}_R, U^{\pm}_R, V^0_R, (V^0)_R^{*}$ are new very heavy right handed bosons. The left and right handed vector bosons are not mixed, and after symmetry breaking the masses of the non-diagonal bosons are given by:
\begin{equation}\label{m.b}
	\begin{array}{c}
	M^{2}_{W_{L,R}}=\frac{1}{4}g^2\left(\nu^{2}_{\eta_{L,R}}+\nu^2_{\rho_{L,R}}\right),\\
	M^{2}_{V_{L,R}}=\frac{1}{4}g^2\left(\nu^{2}_{\eta_{L,R}}+\nu^{2}_{\chi^{'}_{L,R}}\right),\\
	M^{2}_{U_{L,R}}=\frac{1}{4}g^2\left(\nu^{2}_{\rho_{L,R}}+\nu^{2}_{\chi^{'}_{L,R}}\right).
	\end{array}
\end{equation}
In this article, only the mass of the $U_L$ boson is relevant to our results. We will ignore $\nu_{\rho_L}$ since $\nu^{'}_{\chi_L}>>\nu_{\rho_L}$.\footnote{In ref. \cite{Dias:2006}, in a 3-3-1 model without the $SU(3)_R$ group, $\nu^2_{\rho_L}=145.5$ GeV. Adopting this value instead of zero does not change appreciably our results.}

Finally, the mass of the sterile neutrinos are given by the Yukawa sector of the model for leptons, composed of 5-Dimensional effective operators,
{\small
\begin{eqnarray}\label{yuk.lep.3l3r}\nonumber
\mathcal{L}^l_{eff}&=&\frac{h_{ab}^{l}}{\Lambda_D}\left(\bar{\Psi}_{aL}\rho_L\right)\left(\rho^\dag_R\Psi_{bR}\right)+\frac{g^D_{ab}}{\Lambda_D}\left(\bar{\Psi}_{aL}\chi_L\right)\left(\chi^\dag_R\Psi_{bR}\right)+\\ \nonumber
&&{}\frac{y^D_{ab}}{\Lambda_D}\left(\bar{\Psi}_{aL}\eta_L\right)\left(\eta^\dag_R\Psi_{bR}\right)+\frac{g^{M}_{ab}}{\Lambda_{M}}\left[\left(\overline{\left(\Psi_{aL}\right)^c}\chi^*_L\right)\left(\chi^\dag_L\Psi_{bL}\right)\right.+\\ \nonumber
&&{}\left.\left(\overline{\left(\Psi_{aR}\right)^c}\chi^*_R\right)\left(\chi^\dag_R\Psi_{bR}\right)\right]+\frac{y^M_{ab}}{\Lambda_M}\left[\left(\overline{\left(\Psi_{aL}\right)^c}\eta^*_L\right)\left(\eta^\dag_L\Psi_{bL}\right)+\right.\\ \nonumber
&&{}\left.\left(\overline{\left(\Psi_{aR}\right)^c}\eta^*_R\right)\left(\eta^\dag_R\Psi_{bR}\right)\right]+H.c,
\end{eqnarray} }
where $\Lambda_D$ and $\Lambda_M$ denotes energy scales (Dirac and Majorana). $\Lambda_D$ is assumed to be in some grand unification scale, and $\Lambda_M$ in the Planck scale. It is necessary that $\nu_{\eta_R}\approx \nu_{\rho_R}\approx \nu_{\chi^{'}_R}\approx \Lambda_D$ in order to give known particles its correct mass range. The neutrino mass matrix can be written based on this Lagrangian,
\begin{equation}\label{m.m.n}
M_\nu=
	\left(
		\begin{array}{cc}
		M_L & M_D\\
		M_{D}^T & M_R
		\end{array}
	\right),
\end{equation}
where $M_L, M_D$ and $M_R$ are 6x6 block diagonal matrices. The terms in each of these matrices are of order $M_L\sim\nu_L^2$, $M_D\sim\nu_L\nu_R$ and $M_R\sim\nu_R^2$. Since $\nu_R>>\nu_L$, we can neglect the $M_L$ terms and apply the seesaw mechanism to obtain the neutrino mass spectrum after diagonalization of this matrix. We obtain two different mass scales for the neutrinos: The masses of the left and right-handed neutrinos. These are given by:
\begin{eqnarray}\label{m.matrix}
M_{\nu_{L}^{'}}&\approx &-M_D\left(M_R\right)^{-1}M_D^{T},\\
M_{\nu_R}&\approx& M_R.
\end{eqnarray}
For the left handed neutrinos we have,
{\small\begin{equation}\label{m.n.l}
M_{\nu_{L}^{'}}=-\frac{\Lambda_M}{4\Lambda_{D}^2}\left(
	\begin{array}{cc}
	y^D\left(y^M\right)^{-1}\left(y^D\right)^T\nu^2_{\eta_L} & 0\\
	0 & g^D\left(g^M\right)^{-1}\left(g^D\right)^T\nu^2_{\chi^{'}_L}
	\end{array}
\right).
\end{equation}}
Choosing the values of these parameters as $\Lambda_M=10^{19}$GeV, $\Lambda_D=10^{15}$GeV, $\nu_{\eta_L}\sim \mathcal{O} (10 GeV)$ and $\nu_{\chi^{'}_L}\sim\mathcal{O} (TeV)$, the value of the left neutrino masses is,
\begin{eqnarray}
M_{\nu_L}&\sim& y^D\left(y^M\right)^{-1}\left(y^D\right)^T eV,\\
M_{N_L}&\sim& 10  g^D\left(g^M\right)^{-1}\left(g^D\right)^T keV.\label{mnl}
\end{eqnarray} 
So, with coupling constants of order one, the model predicts 3 neutrinos with an eV mass, and 3 sterile neutrinos with keV mass. As mentioned by the authors of \cite{331LR}, there is no mixing between $\nu_L$ and $N_L$, which protects $N_L$ against the X-Ray decay bound, $N_L\rightarrow\nu_L+\gamma$.

\section{Sterile Neutrinos Temperature and Abundance}

In the standard treatment, the value of the ratio between active neutrinos and photon temperature, $T_\nu/T$, is obtained by the conservation of entropy in the primordial plasma\cite{N.Cos,Dodelson}. With the active neutrinos already decoupled, the value of the photon temperature, when compared with $T_\nu$, can be determined by counting the number of degrees of freedom before and after $ee^+$ annihilation, together with the fact that the entropy is proportional to $T^3$. The known result relating neutrino and photon temperatures and abundances can be obtained:
\begin{equation}
T_\nu = \left(\frac{4}{11}\right)^{1/3}T_\gamma ~~~;~~~ n_\nu=\frac{3}{11}n_\gamma
\end{equation}

In the case of sterile neutrinos produced in thermal equilibrium, the argument is similar. The entropy density is given by,
\begin{equation}\label{entropy1}
s_R(T)=g_s(T)\frac{2\pi^2}{45}T^3,
\end{equation}  
where $g_s(T)$ is the {\it entropic degrees of freedom}. $g_s$ counts the number of internal degrees of freedom of the coupled particles. The entropy of the decoupled particles is conserved separately. Since $a\sim T^{-1}$, we have,
\begin{equation}\label{cons.law}
s(T)a^{3}=\text{constant}.
\end{equation}
Suppose that the sterile neutrinos decouple at a temperature $T_D$ with $g_{si}$ entropic degrees of freedom. Right before the active neutrino decoupling, $g_s=2+(7/8)(2.2+3.2)=10.75$ (when $e,e^+$, photons and neutrinos are coupled). So using (\ref{entropy1}) and (\ref{cons.law}), we have:
\begin{equation}\label{tnl/tnu}
g_{si}T_{N_L}^3=10.75T_{\nu}^3\Rightarrow \frac{T_{N_L}}{T_\nu}=\left(\frac{10.75}{g_{si}}\right)^{1/3}.
\end{equation}

As for the abundance, it is important to note that the keV neutrinos today are non-relativistic. In this regime, its energy density today is given by, 
\begin{equation}
\rho_{N_{aL}}(t_0)=n_{N_{aL}}m_{N_{aL}}.
\end{equation}
where $n_{N_{aL}}$ is the number density of the $N_{aL}$ neutrinos and $m_{N_{aL}}$ denotes its mass. With the usual definition of $\Omega_{N_{L}}=\rho_{N_L}/\rho_{cr}$ and $\Omega_{DM}=\rho_{DM}/\rho_{cr}$, we have:
\begin{equation}
\Omega_{N_{L}}h^2=\frac{\sum_{a=e,\mu\tau}n_{N_{aL}}m_{N_{aL}}}{\rho_{cr}}
\end{equation}   
where h is the dimensionless hubble parameter. For simplicity, we make a critical assumption, and consider that all the sterile neutrinos have the same mass. This ensures that there will be no decays from one sterile neutrino to another. So $\sum_{a=e,\mu\tau}n_{N_{aL}}m_{N_{aL}}=n_{N_L}m_{N_L}$. 

Since the number density of a relativistic fermion is roughly given by an integral of the Fermi-Dirac distribution, $f_{FD}$, which is proportional to the cube of its temperature, and all the neutrinos involved decouple before annihilation, it is expected that this should be the case even when they become non-relativistic. That is to say that the ratio between active and sterile neutrinos abundances is actually a ratio between the cube of the temperatures of $N_L$ and $\nu_L$. Introducing such relation in the well-known formula for $\Omega_\nu$, the fraction of the energy density related to neutrinos, and using equation (\ref{tnl/tnu}), we obtain:
\begin{equation}
\Omega_{N_L}h^2
=\frac{m_{N_L}}{94.1\,~{\rm eV}}\left(\frac{T_{N_L}}{T_\nu}\right)^3
=\frac{114.2}{ g_{si}}\left(\frac{m_{N_L}}{1~{\rm keV}}\right)
\end{equation}
By this expression alone is possible to see that keV neutrinos produced 
thermally would close the universe unless they decouple very early. We will get 
back to this point when analyzing the specific model presented here.

Considering that a fraction $\xi$ of dark matter is made of sterile neutrinos, we have:
\begin{equation}
\Omega_{N_L}=\xi\times\Omega_{DM} = 0.26\xi,
\end{equation} 
and using h=0.67\cite{PDG}, we obtain:
\begin{equation}\label{abundance}
\frac{978.5}{g_{si}}\left(\frac{m_{N_L}}{1~{\rm keV}}\right)=\xi
\end{equation}

It is possible to use the above results to constrain the allowed values of the matrix elements $g^D_{ab}$ and $g^M_{ab}$ that appears in the Yukawa Lagrangian. Using the mass of the sterile neutrinos given by  eq. (\ref{m.n.l}), assuming that all of these matrices are proportional to the identity matrix, with values $g_D$ and $g_M$ at the diagonal, and fixing $\Lambda_D$ and $\Lambda_M$ at the GUT and Planck scale respectively, then the neutrino mass sum is given approximately by $m_{N_L}\approx 30\nu^{2}_{\chi^{'}_L}{g_D}^2/4g_M$keV, with $\nu_{\chi^{'}_L}$ in TeV. So, it is possible to relate the values of $g^D$ and $g^M$ with $\xi$ using eq. (\ref{abundance}).

It is important to note that we are always considering the instantaneous decoupling approximation and that products of annihilations are composed of particles still coupled to the plasma. So, in principle, all that is needed is to determine the decoupling temperature of the sterile neutrinos and the entropic degrees of freedom at this temperature.

\section{The sterile neutrinos $N_{eff}$}\label{sect.neff}

We now proceed to the determination of the basic equations for $N_{eff}$ of the sterile neutrinos expected to be found in the early universe. The so called {\it Effective Number of neutrino species}, $N_{eff}$, describes the effect of additional particles in the primordial plasma, in particular, how these new particles change the radiation energy density in the early universe, $\rho_R$. The impact of these particles is measured in terms of the number of neutrinos with standard temperature($T_{\nu}=(4/11)^{1/3}T_\gamma$) that would have an equivalent effect. So, after the $ee+$ annihilation, the energy density of radiation is given by:
\begin{eqnarray}\label{Neff}\nonumber
\rho_R&=&\rho_\gamma+3\rho_\nu+\sum_{i, boson}\rho_i+\sum_{j,{\text fermion}}{\rho_j}\\ 
&=&\rho_\gamma\left[1+\frac{7}{8}\left(\frac{4}{11}\right)^{4/3}N_{eff}\right].
\end{eqnarray}
Since\cite{N.Cos},
\begin{eqnarray}
\rho_R&=&g_{*}\frac{\pi^2}{30}T^4,\\
g_*&=&\sum_{i,\text{boson}}{g_i\left(\frac{T_i}{T}\right)^4}+\frac{7}{8}\sum_{j,{\text fermion}}{g_j\left(\frac{T_ j}{T}\right)^4}.
\end{eqnarray}
By using $T_\nu/T=(4/11)^{1/3}$ and $\rho_\gamma=g_\gamma \pi^2T^4/30$ when necessary, after a few manipulations we finally obtain an expression for $N_{eff}$:
\begin{eqnarray}\label{Neff2}\nonumber
N_{eff}&=&3+\sum_{{\text boson},i}{\frac{g_{i}}{g_\gamma}\left(\frac{8}{7}\right)\left(\frac{11}{4}\right)^{4/3}\left(\frac{T_{i}}{T}\right)^4}+\\
&&{}\sum_{{\text fermion},j}{\frac{g_{j}}{g_\gamma}\left(\frac{11}{4}\right)^{4/3}\left(\frac{T_{j}}{T}\right)^4},
\end{eqnarray}
where $g_{i}$ and $g_{j}$ are the internal degrees of freedom of the bosons and fermions involved. Usually $N_{eff}$ is written as $N_{eff}=3+\Delta N_{eff}$. If there are no new light particles that are relativistic at this time, $N_{eff}=3$.\footnote{Actually, this is the value assuming an instantaneous decoupling of the neutrinos and the plasma. When an non-instantaneous decoupling is considered, $N_{eff}=3.046$.} However, if new light particles are present, the value of $N_{eff}$ will change. This will in turn change the expansion rate of the universe, since in this period the expansion rate is sensitive to $\rho_R$, and can have an impact on cosmological observables, like the Cosmic Microwave Background or Big Bang Nucleosynthesis. 

In the model we are considering, only the keV sterile neutrinos are possible as additional light particles. So the bosonic term desappears, and the fermionic term has only the sterile neutrinos.

The experimental value given to $N_{eff}$ is based on the reference \cite{Cooke}: $N_{eff}=3.28\pm0.28$. We will compare our results with this value.

So, in order to determine the values of $N_{eff}$, it is necessary to know the value of $T_{N_L}/T_\nu$, the ratio between sterile and active neutrino temperatures. This is related to the decoupling temperature of the sterile neutrinos.

Finally, the above result allows us to calculate $N_{eff}$ in expressions (\ref{Neff2}). Using $T_{\nu}=(4/11)^{1/3}T$ when necessary:
\begin{equation}\label{Dneffgs}
\Delta N_{eff}=\left(\frac{10.75}{g_{si}}\right)^{4/3} \text{(per neutrino species),}
\end{equation}

\section{Decoupling temperature of the $N_{aL}$ sterile neutrinos}

One important issue regarding the abundance and $N_{eff}$ of the keV neutrinos in the model is the temperature that they decouple from the primordial plasma. The mechanism behind this fact is quite simple: As the universe expands and gets colder, its temperature falls below the mass of some of the particles in the plasma. When this happens, pair creation of these particles becomes largely supressed, altough pair annihilation can still happen. If these particles are still coupled to the plasma, pair annihilation will destroy them and the liberated entropy will heat the plasma. If, on the other hand, the particle already decoupled, them it does not interact anymore and its temperature goes like $T\sim a^{-1}$ (where $a$ is the scale factor) regardless of what is happening in the plasma.

Active neutrinos have a lower temperature than the CMB photons for this reason: They decouple before the $ee^+$ annihilation, so the photons are heated by this event, and neutrinos are not. But this decoupling, at $T\sim 1 MeV$, happens after other events that have heated the plasma, like $\mu\mu^+$ annihilation, pion annihilation and hadronization. If the sterile neutrinos of the model decouple before any of these events, they will not share the energy liberated by them and, thus, will have smaller temperatures than $T_\nu$, the active neutrino temperature.

A particle is considered coupled to the primordial plasma, if its interaction rate ($\Gamma$) is much larger than the expansion rate of the Universe ($H$), and, at a first approximation, the decoupling temperature $T_D$ is the one that makes the equality $\Gamma(T_D)=H(T_D)$ hold. In the Standard Model, the average interaction rate of the active neutrinos and the electrons in the plasma\footnote{The interaction between electrons and neutrinos, through charged and neutral currents, makes the neutrinos stay in thermal equilibrium with the electrons. Since the electrons are also in equilibrium with the photons, a thermal equilibrium between neutrinos and photons happens.} is given by $\Gamma(T)=G_F^2 T^5$, where $G_F$ is the Fermi constant.

The Fermi constant is related to the W boson mass, that mediates some of the reactions of the active neutrinos: $G_F=\frac{\sqrt{2}}{8}\left(\frac{g}{M_W c^2}\right)^2(\hbar c)^3.$ However, the sterile neutrinos of the 3L3R model do not interact through the W or Z bosons, but only through the $U_L$ and $Z_L^{'}$ bosons, that have a much larger mass. So, in our calculations, we will use an effective Fermi constant, 
\begin{equation}
G_F^{'}=\frac{\sqrt{2}}{8}\left(\frac{g}{M_U c^2}\right)^2(\hbar c)^3=\left(\frac{M_W}{M_U}\right)^2G_F.
\end{equation}

This replacement is made because the $U_L$ now is the mediator of charged currents with the sterile neutrinos $N_{aL}$. Since $M_U>>M_W$, then $G_F^{'}<<G_F$ and the decoupling temperature for $N_{aL}$ should increase in comparison with this temperature for the active neutrinos.

Using $H(T)=\sqrt{\frac{8\pi G}{3}g_*\frac{\pi^2}{30}T^4}\sim \sqrt{g_*}\frac{T^2}{m_{pl}}$ \cite{N.Cos} and making the equality $\Gamma_{N_{aL}}(T_D)=H(T_D)$ we get:

\begin{equation}\label{td}
G_F^{'}T_D^5=\sqrt{g_*}\frac{T_D^2}{m_{pl}}\Rightarrow T_D=\left(\frac{M_U}{M_W}\right)^{4/3}g_*^{1/6} \text{MeV}.
\end{equation}

It is possible to see in equation (\ref{td}) that the decoupling temperature $T_{D}$ of sterile neutrinos is strongly dependent on the $M_U/M_W$ ratio, and weakly dependent on the relativistic degrees of freedom. So, we shall leave the value of $g_*$ constant, and equal to $g_*=16$ (a value that considers that only electrons, positrons, photons and active and sterile neutrinos are fully relativistic at the time of decoupling), and adopt different possible values of $M_U$.

\section{Results}

In this final section, we want to give the values of $N_{eff}$ and the allowed values of $g^D$ and $g^M$ for three different cases. (1) When $N_{aL}$ decouple between pion and muon annihilation $(T\in[105,140 MeV])$, (2) when the decoupling happens between hadronization and pion annihilation $(T\in[140,200 MeV])$, (3) when $N_{aL}$ decouples before hadronization (in the interval $T\in[200,220]$ MeV). We adopt $T_{Had}=200 MeV$\cite{N.Cos}.

Given our assumptions, the decoupling temperature $T_D$ given in (\ref{td}) will be in case (1) if $M_U/M_W\in[23.2,28.8]$, in case (2) if $M_U/M_W\in[28.8,37.6]$ and in case (3) if, approximately, $M_U/M_W\in[37.6,40]$. Relating these mass ranges with equation (\ref{m.b}) for $M_{U_L}$, the allowed values for $\nu_{\chi_L^{'}}$ is $\nu_{\chi_L^{'}}\in[5.7,7.2]$ TeV in case (1), $\nu_{\chi_L^{'}}\in[7.2,9.3]$ TeV in case (2) and $\nu_{\chi_L^{'}}\in[9.3,9.9]$ TeV in case (3).

So, we have\footnote{To our knowledge, there are no experimental limits on the mass of the $U_L$ bosons of this model. There are experimental limits on the mass of the $Z^{'}$ (the $Z_L^{'}$ in the 3L3R) of a 3-3-1RH model\cite{Dias:2006}, which is similar to the left-handed sector of the 3L3R. These limits are given in\cite{z'mass}. The limit given is $M_{Z^{'}}\geq 2.2 TeV$. This implies that $M_U/M_W\geq 25$. The results given in \cite{z'mass} are not completely applicable to the 3L3R. But if they are used (to give at least an estimate of a lower limit of the $U_L$ mass) $N_{aL}$ never decouples before muon annihilation.}:

\bigskip
Case (1):

\smallskip
In this case, only muons, anti-muons, electrons, positrons, neutrinos and photons are coupled. This gives,  
\begin{equation}
g_{si}=\underbrace{2}_{\gamma}+\frac{7}{8}(\underbrace{2.2}_{e,e^+}+\underbrace{2.2}_{\mu,\mu^+}+\underbrace{3.2}_{\nu_L,\bar{\nu}_L})=14.25,
\end{equation}
which implies, from (\ref{Dneffgs}):
\begin{equation}
\Delta N_{eff}=0.69.
\end{equation}

Given that the experimental limit is $\Delta N_{eff}=0.28\pm0.28$, even one additional sterile neutrino would be excluded by $1\sigma$.
\bigskip

Case (2):

\smallskip
In this case, pions(neutral and charged) are now coupled, and we have,
\begin{equation}
g_{si}=\underbrace{2}_{\gamma}+\underbrace{3}_{\pi}+\frac{7}{8}(\underbrace{2.2}_{e,e^+}+\underbrace{2.2}_{\mu,\mu^+}+\underbrace{3.2}_{\nu_L,\bar{\nu}_L})=17.25,
\end{equation}
which gives,
\begin{equation}
\Delta N_{eff}=0.53.
\end{equation}

Now $N_{eff}$ allows only one additional sterile neutrino at the $1\sigma$ level.

\bigskip
Case (3):
\smallskip

Before the hadronization, gluons and the quarks up, down and strange, and its respective antiquarks, are coupled to the plasma. This implies in a huge increase in $g_{si}$, as it is shown below:
{\small
\begin{equation}
g_{si}=\underbrace{2}_{\gamma}+\underbrace{8.2}_{\text{gluons}}+\frac{7}{8}(\underbrace{2.2}_{e,e^+}+\underbrace{2.2}_{\mu,\mu^+}+\underbrace{3.2}_{\nu_L,\bar{\nu}_L}+\underbrace{3.3.2.2}_{u,\bar{u},d,\bar{d},s,\bar{s}})=61.75,
\end{equation}}
which implies,
\begin{equation}
\Delta N_{eff}=0.097.
\end{equation}

Now three additional neutrino species are allowed, with a total $\Delta N_{eff}=0.29$ in this case.

Limit Case:
\smallskip

We could think of a scenario where $g_{si}$ assumes the maximum possible value for the 3L3R, a limit case, that should maximize the allowed region for $g_M$ and $g_D$. This could happen if these neutrinos decoupled before every other particle in the left sector of the model. That is to say, it should decouple before 9 quarks, 6 leptons, 10 scalar bosons and 17 gauge bosons, with $g_{si}=162.25$ and $\Delta N_{eff}$ would allow three sterile neutrinos. Unfortunately, the required value for $\nu_{\chi^{'}_L}$ in this case should also be large, and would more than compensate the larger value of $g_{si}$, giving a very restricted parameter space for $g_D$ and $g_M$. So we ignore this limit case.

The abundance arguments gives the following results for $g_D$ and $g_M$ parameter space, as shown in Figure 1.
	
Figure 1 makes clear that values for which $N_L$ have masses in the keV range are completely excluded, by orders of magnitude. Indeed, the allowed mass range for these neutrinos lie in the (in most cases) low eV range, as shown in table 1.

\begin{figure}[h!]
\begin{minipage}{0.48\linewidth}
\includegraphics[width=\linewidth]{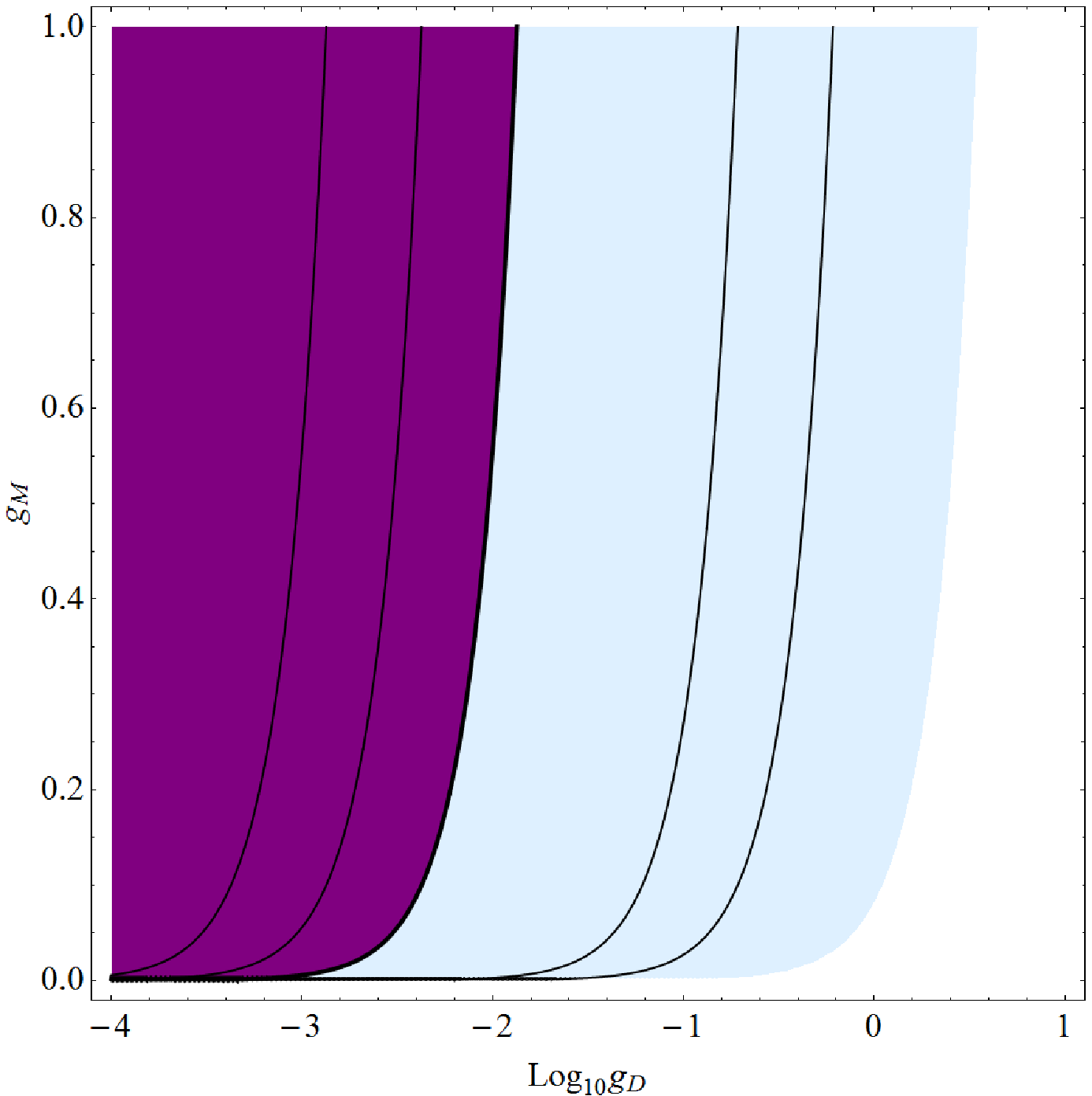}
\end{minipage}\hfill
\begin{minipage}{0.48\linewidth}
\includegraphics[width=\linewidth]{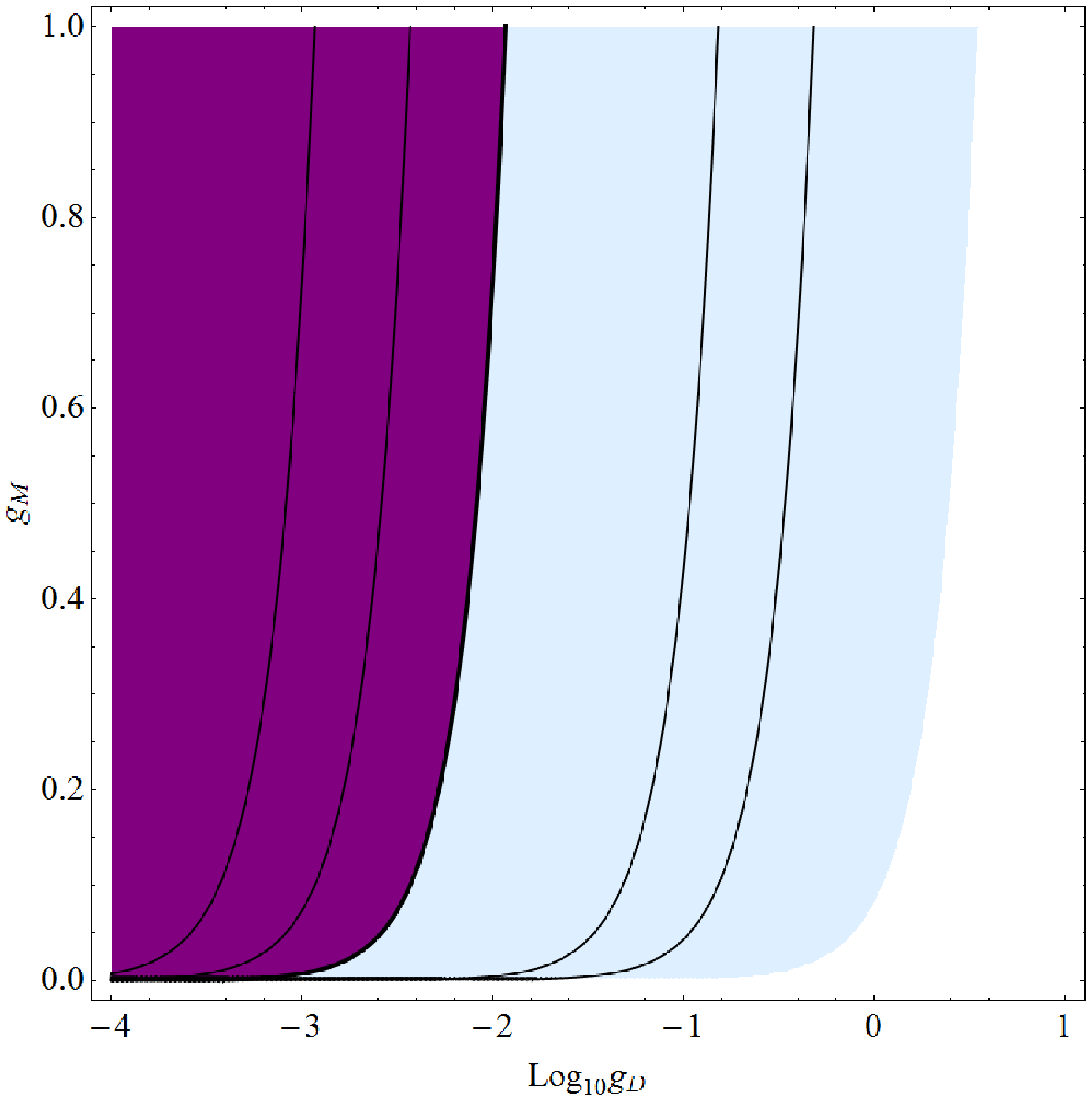}
\end{minipage}
\begin{minipage}{0.48\linewidth}
\includegraphics[width=\linewidth]{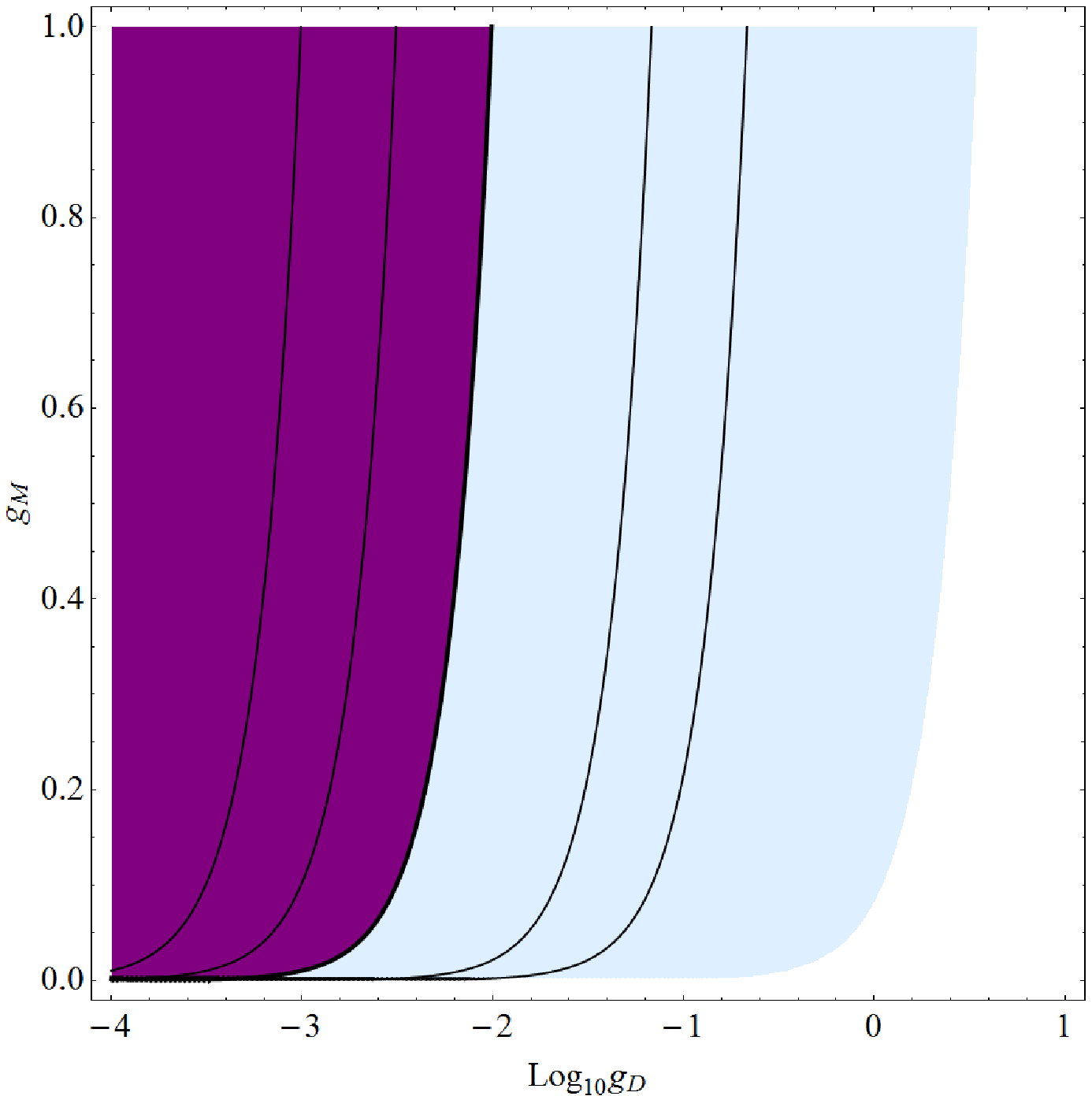}
\end{minipage}\hfill
\caption{{\footnotesize Graphics of the allowed regions of parameter space for $g_D$ and $g_M$. Upper left: Case 1 ($\nu_{\chi^{'}_L}=5.7$ TeV). Upper right: Case 2 ($\nu_{\chi^{'}_L}=7.2$ TeV). Lower figure: Case 3 ($\nu_{\chi^{'}_L}=9.3$ TeV). Cases 1 and 2 have only one neutrino, and case 3 has three. In each graphic, from left to right, the first three isolines gives the values of $g_D$ and $g_M$ for $\xi=0.01$, $\xi=0.1$ and $\xi=1$, with the thick line representing $\xi=1$. The last two isolines gives values for which the sterile neutrino mass is $m_{N_L}=1$ keV and $m_{N_L}=10$ keV, respectively. The dark purple area denotes the region consistent with the constraint $\xi\leq1$. It is clear from this figure that keV masses are not allowed.}}
\end{figure}

Dark Matter cannot be made only of these light particles. For example, limits based on dwarf spheroidal galaxies and lyman-$\alpha$ forest, gives a lower bound of $M_N>O(1)$ keV if all of Dark Matter is made sterile neutrinos\cite{DarkMatter}, which excludes the $\xi=1$ scenario. Since they are Hot Dark Matter (HDM), a more realistic constraint on the values of $g_D$ and $g_M$ must assume low values of $\xi$. For instance, $\xi\leq0.1$ and $\xi\leq0.01$ in Figure 1 would be a safer region, since structure formation allows that $\Omega_{HDM}$ is at most only a few percent of $\Omega_{DM}$. However, since only a very small fraction of DM can be made of these neutrinos, the 3L3R is unable to solve the DM puzzle.

Table 2 shows the maximum values that $g_D$ can have for each case, depending on the value of  $\nu_{\chi_L^{'}}$. Case 1 is a little better than case 2 regarding allowed values for $g_D$. However, as we've seen, this case does not satisfy the $N_{eff}$ constraint and is therefore excluded. Case 2 is allowed, but in strong tension with $N_{eff}$ (since only one neutrino species is permitted, and at the very edge of 1 sigma) and at odds with our calculations that assumes 3 new neutrinos. Case 3 satisfies $N_{eff}$ and its allowed parameter space region is given in table 2.

\begin{table*}[t!]
\caption{Values of the allowed neutrinos mass for each scenario and neutrino energy fraction of Dark Matter($\xi$). It is clear that in all cases, the sterile neutrino masses must be in the eV-range or lower in order to satisfy the constraint $\xi\leq1$.}
\begin{ruledtabular}
\begin{tabular}{cccccc}
	\multicolumn{6}{c}{Cases}\\ \hline
	\multicolumn{2}{c}{Case 1 (one neutrino)} & \multicolumn{2}{c}{Case 2 (one neutrino)} & \multicolumn{2}{c}{Case 3 (three neutrinos)} \\\hline
	$\xi$ & $m_{N_{aL}}$(eV)&$\xi$ & $m_{N_{aL}}$(eV)&$\xi$ & $m_{N_{aL}}$(eV)\\ 
	0.01 & 0.14 & 0.01 & 0.17 & 0.01 & 0.21\\ 
	0.1  & 1.45 & 0.1  & 1.76 & 0.1  & 2.1 \\ 
	1  & 14.56 & 1  & 17.62 & 1  & 21.03 \\ 
	68.66  & 1000 & 56.62  & 1000 & 47.53  & 1000 \\ 
	686.67 & 10000 & 567.25 & 10000 & 475.38 & 10000 \\ 
	\end{tabular}
	\end{ruledtabular}
	\end{table*}

\begin{table*}[t!]
\caption{Maximum allowed value for $g_D$ in each scenario, with $g_M=1$ and $\nu_{\chi^{'}_L}=5.7$ TeV (case 1), $\nu_{\chi^{'}_L}=7.2$ TeV (case 2) and $\nu_{\chi^{'}_L}=9.3$ TeV (case 3). A greater value for $\nu_{\chi_L^{'}}$ implies a lower allowed value for $g_D$, so the maximum value for $g_D$ in each case corresponds to the minimum value of  $\nu_{\chi_L^{'}}$ in that case.}
\begin{ruledtabular}
\begin{tabular}{cccc}
	\multicolumn{4}{c}{Cases}\\ \hline
	$\xi$ & Case 1 (one neutrino) & Case 2 (one neutrino) & Case 3 (three neutrinos) \\\hline 
	0.01 & $1.3\times 10^{-3}$  & $1.1\times 10^{-3}$  & $9.8\times10^{-4}$ \\
	0.1 & $4.2\times10^{-3}$ & $3.6\times10^{-3}$ & $3.1\times10^{-3}$
	\end{tabular}
	\end{ruledtabular}
	\end{table*}

Since these candidates are hot dark matter, another different limit can be imposed. By reference \cite{PDG}, the sum of neutrino masses by cosmology is $\sum m_\nu <0.23$ eV (results from Planck + BAO). The sterile neutrinos of the 3L3R should have an impact in this result, since they are hot dark matter just as the active neutrinos. Indeed, using the $\Omega_\nu$ relation again, it is possible to deduce that the sterile neutrinos affect the neutrino sum masses as $\sum_a[m_{a\nu}+(n_{N_{aL}}/n_\nu)m_{N_{aL}}]<0.23$ eV. Since the minimum sum of active neutrino masses by oscillation experiments is $\sum m_\nu \geq0.06$ eV, we have:
\begin{equation}\label{s.n.sum}
\sum_a (n_{N_{aL}}/n_\nu)m_{N_{aL}}\leq 0.17 eV.
\end{equation}

For case 2, $(n_{N_{aL}}/n_\nu)=0.62$ and allows only one neutrino, and case 3 has $(n_{N_{aL}}/n_\nu)=0.17$ and 3 neutrinos. For $\xi=0.01$, we have for case (2) $\sum_a (n_{N_{aL}}/n_\nu)m_{N_{aL}}= 0.62\times 0.17 \approx0.11$ eV and case (3) $\sum_a (n_{N_{aL}}/n_\nu)m_{N_{aL}}= 0.17\times 3\times0.21 \approx 0.11$ eV. So for both cases the bound given in (\ref{s.n.sum}) is satisfied. 

As expected by the value of allowed masses, note that the $\xi=0.1$ scenario is excluded, since relation (\ref{s.n.sum}) is not satisfied.

With corrections given by the fact that only allows one neutrino species, case 2 has a little larger parameter space than case 3. Both cases obey the cosmological bounds applied and, although unable to answer the DM problem, are not ruled out as HDM candidates.

\section{Conclusion}

In this paper we analysed the feasability of sterile neutrinos in an extension of the SM, called 3L3R extension, as candidates for warm dark matter. These neutrinos arise naturally in this model, with a mass $\sim$ keV, and at first seem to satisfy the requirements for dark matter. Since these neutrinos interact through gauge bosons with SM particles, it is expected that they were in thermal equilibrium with them some time in the past, allowing us to determine their number density as an integral of the Fermi-Dirac distribution, just like it is done with active neutrinos.

We imposed the $N_{eff}$ and abundance constraints on these particles, and analysed them in 3 different cases. It was shown that the $N_{eff}=3.28\pm0.28$ constraint was satisfied in 2 of the 3 cases, and case 1 is excluded at the $1\sigma$ level. In case (2) only one additional neutrino was allowed and $\Delta N_{eff}=0.53$ and in case (3) all three neutrinos were possible, giving a total $\Delta N_{eff}=0.29$. The decoupling temperature in each case also gives lower bounds to the $U_L$ boson mass, with $M_{U_L}\sim2.3-3$ TeV in case (2) and $M_{U_L}> 3 $ TeV in case (3).

The thermalized nature of these neutrinos, together with the assumption that they have roughly equal masses and are not CDM, implies that they  must be much lighter than a keV mass and are Hot Dark Matter particles. Being HDM they can only constitute a fraction of DM. This can give some constraints on the allowed mass, shown in table 1 (of order 0.1 eV for $\Omega_N\sim0.01 \Omega_{CDM}$ and $~2$ eV for $\Omega_N\sim0.1 \Omega_{CDM}$) and on the allowed values for the $g_D$ and $g_M$ parameters in the Yukawa lagrangian of the model. This analysis have been made, and the allowed regions for these parameters were shown in Figure 1. If at most $1\%$ of DM is made of the 3L3R, then $g_D\leq1.1\times10^{-3}$ (case 2) and $g_D\leq9.8\times10^{-4}$ (case 3).

The HDM candidates would impact $\sum m_\nu$. Combining the minimal active neutrino sum mass by terrestrial experiments ($\sum m_\nu\geq0.06$ eV) and the cosmological limit on these masses ($\sum m_\nu\leq0.23$ eV), it is possible to deduce that $\sum_a (n_{N_{aL}}/n_\nu)m_{N_{aL}}\leq 0.17$ eV. This relation is satisfied in cases (2) and (3), with a value of $\sim0.11$ eV for $\Omega_{N_{L}}=0.01\Omega_{CDM}$. If its energy density is equal to $10\%$ of CDM, then the above relation is not satisfied. 

It is also important to note that these conclusions of overabudance of keV particles are applicable to any model that predicts stable particles in thermal equilibrium with the primordial plasma, through the exchange of gauge bosons, in the absence of mixing. Finally, the above results are a consequence of the fact that the neutrinos in the 3L3R are stable, and do not mix with the active neutrinos. If this were not the case, the results could change. For example, if the neutrinos have different masses, with one of the neutrinos heavy enough and relatively long lived, it could decay out of equilibrium into another sterile neutrino, producing a large amount of entropy that dilutes the abundance of the remaining, stable neutrinos. This could produce the correct abundance with keV sterile neutrinos, a result shown in ref. \cite{Bezrukov}. We can apply this mechanism to the 3L3R in the future.

So, given our assumptions, we conclude that only cases 2, 3 are consistent with the $N_{eff}$ constraint, and, even in these cases, the allowed values for the sterile neutrino masses and coupling constants in the Yukawa sector is very small, since the abundance constraint implies a very low mass. These neutrino are viable HDM particles.

\acknowledgments

The authors are grateful for the Conselho Nacional de Desenvolvimento Científico e Tecnológico (CNPq) and the Fundação de Amparo à Pesquisa do Estado de São Paulo (FAPESP) for the financial support.

\end{document}